\documentclass[letter]{jpsj2} 
\usepackage{bm}
\usepackage{graphicx}
\newcommand{\bmk}{{\bm k}}
\newcommand{\bmp}{{\bm p}}
\newcommand{\bmq}{{\bm q}}
\newcommand{\sgn}{{\rm sgn}}
\newcommand{\Ln}{{\rm Ln}}
\newcommand{\Arg}{{\rm Arg}}
\renewcommand{\Im}{{\rm Im}}
\newcommand{\tildekf}{{\widetilde k}_F}

\title{
 Perturbative calculation of the spin-wave dispersion in a
disordered double-exchange model
}

\author{Taeko Semba and Takahiro Fukui}

\inst{Department of Mathematical Sciences, 
Ibaraki University, Mito 310-8512, Japan}

\recdate{\today}

\abst{We study the spin-wave dispersion of localized spins in 
a disordered double-exchange model using the perturbation theory
with respect to the strength of the disorder potential.
We calculate the dispersion upto the next-leading order, and 
extensively examine the case of one-dimension. 
We show that in that case, disorder  yields anomalous
gapped-like behavior at the Fermi wavenumber of the conduction
electrons.

}

\kword{colossal magnetoresistance manganites, double-exchange model, 
randomness, spin-wave dispersion, perturbation}

\begin{document}
\maketitle




Colossal magnetoresistance manganites \cite{HWH,COK,JTM}
have been providing hot topics owing to their rich magnetic and 
electronic properties.\cite{IFT,SalJai,DHM}
Recent studies have clarified that the 
double-exchange (DE) mechanism \cite{Zen,AndHas,Gen,KubOha} alone
is not enough to understand a variety of phases of these materials,
which is actually induced by the interplay between spin, charge and orbital
degrees of freedom.

Recently, the softening of the spin-wave dispersion,
\cite{HDC,DLM,DHZ,BHM,AUT}
observed in R$_{1-x}$D$_x$MnO$_3$ 
(R$=$La, Pr, Nd and D$=$Ba, Sr, Ca, etc.),
has suggested that a random potential also plays a crucial role 
in magnetic properties of these materials.
\cite{Var,MueDag,SXS1,SXS2,AllAla,BMM}
Although some other mechanism 
\cite{Fur99,KhaKil,SolTer,Gol,ShaChu}
have been proposed to describe the
softening phenomena, 
recent numerical calculations by Motome and Furukawa
\cite{MotFur1,MotFur2,MotFur3} have shown that 
a simple DE model with a random potential can
explain the anomalous behavior of the spin-wave
spectrum including broadening, gap formation as well as softening.

Motivated by their work, we study a disordered DE model
by the use of the perturbation with respect to the strength of
a random potential.
Based on a general formula we derived, 
we show that in the case of 1-dimensional (1D) systems
randomness yields a singularity of the spin-wave dispersion
at the Fermi wavenumber of conduction electrons. 
It turns out that such anomalous behavior is mainly due to the nesting of
the Fermi surface in 1D models.
We then discuss the higher dimensional systems,
based on the results of 1D systems.

We start with the following Hamiltonian:
\begin{eqnarray}
H=-t\sum_{\langle i,j\rangle}\sum_\sigma 
c_{i\sigma}^\dagger c_{j\sigma}
-J_H\sum_i \bm{s}_i\cdot\bm{S}_i 
+\sum_i\epsilon_ic_{i\sigma}^\dagger c_{i\sigma},
\end{eqnarray}
where 
$\bm{s}_i
=\frac{1}{2}c_{i\sigma}^\dagger\bm{\sigma}_{\sigma\sigma'}c_{i\sigma'}$
and $\bm{S}_i$
denotes, respectively,  
the electron spin and the localized spin in the spin-$S$ representation, and
$\epsilon_i$ is the quenched random potential
with statistical properties $\overline{\epsilon_j}=0$ and 
$\overline{\epsilon_i\epsilon_j}=g\delta_{i,j}$.
This model includes two large parameters, $S$ and $J_H$.
Assuming that localized spins are almost aligned to the $z$-direction,
we first take into account the leading terms with respect to 
the series of $1/S$.
To this end, we utilize the Holstein-Primakoff mapping for localized spins:
\begin{eqnarray}
S_i^+=\sqrt{2S}a_i,
\quad
S_i^-=\sqrt{2S}a_i^\dagger,
\quad
S_i^z=S-a_i^\dagger a_i ,
\end{eqnarray}
which leads us to the following truncated Hamiltonian up to $O(1/S^0)$:
\begin{eqnarray}
H&=&-t\sum_{\langle i,j\rangle, \sigma}
c_{i\sigma}^\dagger c_{j\sigma}
-\frac{J_HS}{2}\sum_{i, \sigma} \sigma c_{i\sigma}^\dagger c_{i\sigma}
\nonumber\\
&&\mbox{}
-\sqrt{\frac{S}{2}}J_H\sum_{i}
(a_i^\dagger c_{i\uparrow}^\dagger c_{i\downarrow}+
a_ic_{i\downarrow}^\dagger c_{i\uparrow})
+\frac{J_H}{2}\sum_{i, \sigma}
a_i^\dagger a_i \sigma c_{i\sigma}^\dagger c_{i\sigma}
+\sum_{i, \sigma}\epsilon_ic_{i\sigma}^\dagger c_{i\sigma}.
\label{StaHam}
\end{eqnarray}
Based on this Hamiltonian,
we calculate the spin-wave dispersion perturbatively with respect to 
the strength of disorder $g$,
collecting the leading terms with respect to $J_H$.

\begin{figure}[hbt]
\begin{center}
\includegraphics[width=0.5\linewidth]{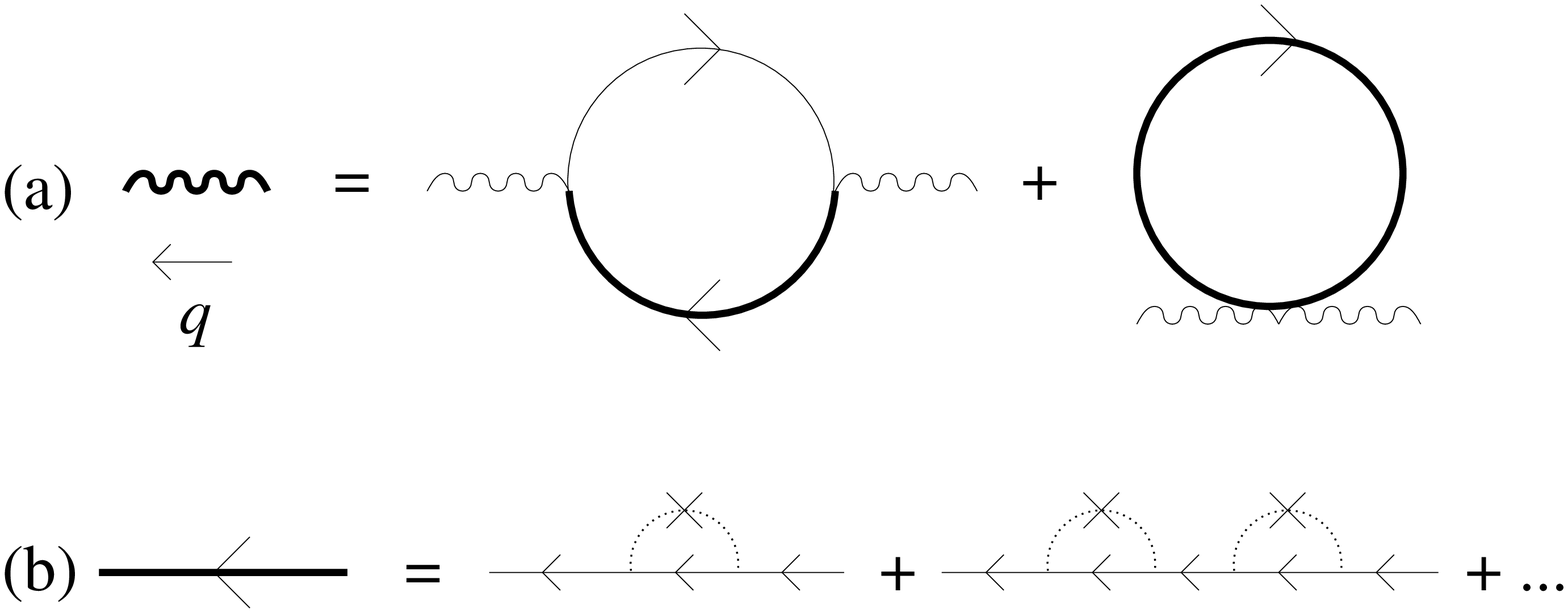}
\caption{\label{fig:LedSelEne}
(a) The leading self-energy of the spin-wave.
Thin-lines denotes the down-spin propagator, whereas the thick-line
denotes the up-spin propagator with the self-energy induced by disorder.
(b) The propagator of up-spin electrons used in the diagrams (a).
The thin-line and dotted-line denotes the bare up-spin propagator and
impurity potential, respectively.}
\end{center}
\end{figure}

Previously, Furukawa \cite{Fur96} calculated the magnon dispersion 
in the clean DE model, according to the diagrams (a)
in Fig.\ref{fig:LedSelEne}.
In the presence of disorder, these diagrams should be evaluated 
by the use of the electron propagator with the self-energy 
induced by the scattering of electrons by impurities. 
Since down-spin electrons are located in the momentum space
far above the Fermi energy in the large $J_H$ limit, 
it is readily seen that disorder yields
the self-energy to up-spin electrons only, which is evaluated as
\begin{eqnarray}
\Sigma_{\uparrow}(i\varepsilon_n)=
\sum_{\bmk}\frac{-g}{i\varepsilon_n-\varepsilon_\bmk}
\sim\frac{i}{\tau}\sgn(\varepsilon_n) 
\end{eqnarray}
in a weak disorder limit,
where 
$\varepsilon_n=(2n+1)\pi/\beta$ 
is the fermionic Matsubara frequency,
$\varepsilon_\bmk=-2t\sum_\mu\cos k_\mu-\varepsilon_F$ 
is the bare electron dispersion, and
the relaxation time $\tau$ is given by 
$1/\tau=\pi g D(\varepsilon_F)$.

\noindent
Using the propagator with the self-energy above, 
the diagrams in Fig.\ref{fig:LedSelEne} (a) are calculated as
\begin{eqnarray}
\Pi_\bmq^{(0)}(i\omega_n)&=&
\frac{J_H^2S}{2}\sum_{i\varepsilon_n}\sum_{\bmk}
\frac{-1}{i\varepsilon_n-\varepsilon_\bmk+\Sigma_\uparrow(i\varepsilon_n)}
\frac{-1}{i(\varepsilon_n+\omega_n)-\varepsilon_{\bmk+\bmq}-J_HS}
\nonumber\\
&&\mbox{}
-\frac{J_H}{2}\sum_{i\varepsilon_n}\sum_\bmk
\frac{-1}{i\varepsilon_n-\varepsilon_\bmk+\Sigma_\uparrow(i\varepsilon_n)}
\nonumber\\
&\rightarrow&
\frac{1}{2S}\sum_\bmk 
f_\tau(\varepsilon_\bmk)(\varepsilon_{\bmk+\bmq}-\varepsilon_\bmk),
\label{LedSelEne}
\end{eqnarray}
where
\begin{eqnarray}
f_\tau(\varepsilon)=
\frac{1}{2}\left(1-\frac{2}{\pi}\arctan\tau\varepsilon\right),
\label{SmeDisFn}
\end{eqnarray}
and the arrow in eq.(\ref{LedSelEne}) implies the large $J_H$ limit.
This equation indicates that disorder merely smears the Fermi surface at
the leading order and hence, does not give rise to anomalous behavior
for the magnon spectrum.

In order to calculate the self-energy of the spin-wave 
at the next leading order, 
let us consider the diagrams in Fig.\ref{fig:NexBarSelEne}.
Each diagram is evaluated as follows:

\begin{figure}[htb]
\begin{center}
\includegraphics[width=0.7\linewidth]{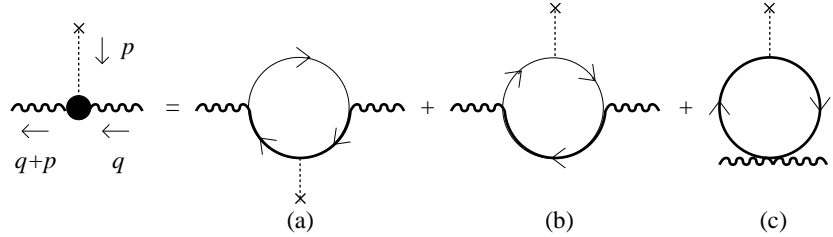}
\caption{\label{fig:NexBarSelEne} Diagrams which contribute to the 
next leading order of the spin-wave self-energy.
The dotted-line denotes the impurity
potential. The thick-wavy-line stands for the spin-wave propagator
in Fig.\ref{fig:LedSelEne}.}
\end{center}
\end{figure}

\begin{eqnarray}
\Pi^{(1a)}_{\bmq+\bmp,\bmq}(i\omega_n;\epsilon_\bmp)
&=&\frac{J_H^2S}{2}\epsilon_\bmp\sum_{\varepsilon_n}\sum_\bmk
\frac{-1}{i\varepsilon_n-\varepsilon_\bmk+\Sigma_\uparrow(i\varepsilon_n)}
\frac{-1}{i\varepsilon_n-\varepsilon_{\bmk+\bmp}+\Sigma_\uparrow(i\varepsilon_n)}
\nonumber\\
&&\mbox{}\times
\frac{-1}{i(\varepsilon_n+\omega_n)
-\varepsilon_{\bmk+\bmp+\bmq}-J_HS}
\nonumber\\
&\rightarrow&
\frac{J_H}{2}\epsilon_\bmp\chi_\bmp
+\frac{1}{2S}\epsilon_\bmp\sum_\bmk f_\tau(\varepsilon_\bmk)
\frac{2i\omega_n+2\varepsilon_\bmk-\varepsilon_{\bmk+\bmp+\bmq}
-\varepsilon_{\bmk-\bmq}}{\varepsilon_\bmk-\varepsilon_{\bmk+\bmp}} ,
\end{eqnarray}
\begin{eqnarray}
\Pi^{(1b)}_{\bmq+\bmp,\bmq}(i\omega_n;\epsilon_\bmp)
&=&\frac{J_H^2S}{2}\epsilon_\bmp\sum_{\varepsilon_n}\sum_\bmk
\frac{-1}{i\varepsilon_n-\varepsilon_\bmk+\Sigma_\uparrow(i\varepsilon_n)}
\frac{-1}{i(\varepsilon_n+\omega_n)
-\varepsilon_{\bmk+\bmq}-J_HS}
\nonumber\\
&&\mbox{}\times
\frac{-1}{i(\varepsilon_n+\omega_n)
-\varepsilon_{\bmk+\bmp+\bmq}-J_HS}
\nonumber\\
&\rightarrow&
-\frac{1}{2S}\epsilon_\bmp\sum_\bmk f_\tau(\varepsilon_\bmk) ,
\end{eqnarray}
and
\begin{eqnarray}
\Pi^{(1c)}_{\bmq+\bmp,\bmq}(i\omega_n;\epsilon_\bmp)
&=&-\frac{J_H}{2}\epsilon_\bmp\sum_{\varepsilon_n}\sum_\bmk
\frac{-1}{i\varepsilon_n-\varepsilon_\bmk+\Sigma_\uparrow(i\varepsilon_n)}
\frac{-1}{i\varepsilon_n-\varepsilon_{\bmk+\bmp}+\Sigma_\uparrow(i\varepsilon_n)}
\nonumber\\
&\rightarrow&
-\frac{J_H}{2}\epsilon_\bmp \chi_\bmp ,
\end{eqnarray}
where $\epsilon_\bmp$ stands for the impurity potential in eq.(\ref{StaHam})
in the momentum representation, $\chi_\bmp$ is defined by 
\begin{eqnarray}
\chi_\bmp=\sum_\bmk
\frac{f_\tau(\varepsilon_\bmk)-f_\tau(\varepsilon_{\bmk+\bmp})}
{\varepsilon_\bmk-\varepsilon_{\bmk+\bmp}} ,
\label{Chi}
\end{eqnarray}
and $\omega_n$ is the bosonic Matsubara
frequency $\omega_n=2n\pi/\beta$.
Collecting these diagrams, we obtain
\begin{eqnarray}
\Pi^{(1)}_{\bmq+\bmp,\bmq}(i\omega_n;\epsilon_\bmp)
=\frac{\epsilon_\bmp}{2S}
\sum_\bmk f_\tau(\varepsilon_\bmk)
\frac{2i\omega_n+\varepsilon_\bmk-\varepsilon_{\bmk-\bmq}
+\varepsilon_{\bmk+\bmp}-\varepsilon_{\bmk+\bmp+\bmq}}
{\varepsilon_\bmk-\varepsilon_{\bmk+\bmp}} .
\end{eqnarray}

\begin{figure}[htb]
\begin{center}
\includegraphics[width=0.3\linewidth]{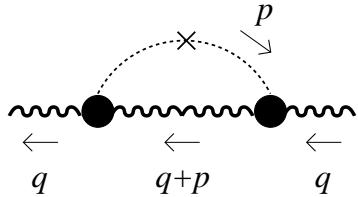}
\caption{\label{fig:NexSelEne}The next-leading self-energy of the spin-wave.}
\end{center}
\end{figure}

As already claimed, the contractions of the impurity lines in the same
electron loop in Fig.\ref{fig:NexBarSelEne},
which corresponds to Fig.\ref{fig:LedSelEne},
do not give any anomalies
in the magnon spectrum. Therefore, one can expect singular behavior
in another type of diagrams, in which impurity lines are contracted 
among  different electron loops in such a way as induces the 
correlations among up-spin electrons through the disorder potential.
One of such simplest diagrams is
given in  Fig.\ref{fig:NexSelEne}. 
To calculate this diagram explicitly,
we use the quadratic dispersion 
$\varepsilon_k\sim t\bmk^2-\varepsilon_F$. 
In 1D systems, this approximation works
basically without loss of generality,
whereas in the case of higher dimensional systems,
it is well-known that 
the above self-energy, the function $\chi_\bmp$ in particular, 
is quite sensible to the form of the dispersion 
relation of bare electrons.
Therefore, we restrict our calculations to 1D systems to reveal 
the mechanism of gap opening in a disordered DE model.
As to higher dimensional systems, 
we will discuss what we expect at the end of the calculation
based on the results of the 1D model.

The use of the quadratic dispersion leads to the following 
expression of the self-energy in Fig.\ref{fig:NexSelEne} 
for the 1D system,
\begin{eqnarray}
\Pi_{q}^{(1)}(i\omega_n)=\frac{gt^2}{S^2}\sum_p
\frac{\left(pq+q^2\right)^2\chi_p^2}{i\omega_n-\Pi_{q+p}^{(0)}}.
\end{eqnarray}
Here, we have neglected the Matsubara frequency in the 
numerator, since it merely gives rise to small effects.
The total self-energy is now given by 
$\Pi_q^{(0)}(i\omega_n)+\Pi_q^{(1)}(i\omega_n)$,
and the spin-wave dispersion relation can be determined by solving
$\omega=\Pi_q^{(0)}(\omega)+\Pi_q^{(1)}(\omega)$.
In what follows, we solve this nonlinear equation iteratively.

At the first step, one finds the spectrum given by 
\begin{eqnarray}
\omega_q=\Pi_q^{(0)}= 2t' (1-\cos q) ,
\end{eqnarray}
where $t'=t/(2S)\sum_k f_\tau(\varepsilon_k)\cos k$.
Next, substituting this 
into $\Pi_q^{(1)}(\omega)$,
we have
\begin{eqnarray}
\omega_q&=&\Pi_q^{(0)}+\frac{gt^2}{S^2}\sum_p
\frac{\left(pq+q^2\right)^2\chi_p^2}{\Pi_q^{(0)}-\Pi_{q+p}^{(0)}}.
\nonumber\\
&\sim&
\Pi_q^{(0)}-\frac{gt^2}{S^2t'}\sum_p
\frac{\left[(pq)^2+q^4\right]\chi_p^2}{p^2+2pq},
\label{NexEq}
\end{eqnarray}
where we have used the quadratic dispersion approximation for
the spin-wave, $\Pi_q^{(0)}\sim t'q^2$.

The second term above should exhibit singular behavior 
in the spin-wave spectrum. 
To observe this, we first calculate the function $\chi_p$ with 
the smeared distribution function $f_\tau(\varepsilon)$
in eq.(\ref{SmeDisFn}). 
We obtain
\begin{eqnarray}
\chi_p=-\frac{1}{2\pi t}\frac{1}{2p}
\ln\left[
\frac{(\frac{p}{2})^2+p\tildekf\cos\frac{\theta}{2}+\tildekf^2}
{(\frac{p}{2})^2-p\tildekf\cos\frac{\theta}{2}+\tildekf^2}
\right] ,
\label{ChiCon}
\end{eqnarray}
where $\tildekf$ and $\theta$ is defined, respectively,
$\tildekf=(k_F^4+\epsilon^2)^{1/4}$ with $\epsilon=1/(t\tau)$ and
$\tan\theta=\epsilon/k_F^2$.
It may be instructive to take the zero disorder limit,
$g\rightarrow0~(\epsilon\rightarrow 0)$ in this function: 
It then becomes the well-known form,
\begin{eqnarray}
\chi_p\rightarrow 
-\frac{1}{2\pi t}\frac{1}{p}\ln\left|\frac{p+2 k_F}{p-2k_F}\right|.
\label{ChiLim}
\end{eqnarray}
\begin{figure}[htb]
\begin{center}
\includegraphics[width=0.5\linewidth]{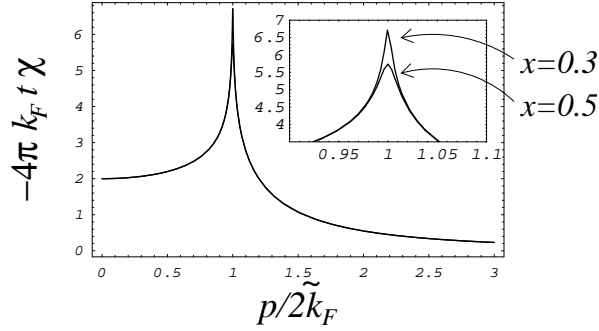}
\caption{\label{fig:Chi} 
The function $\chi$ in eq.(\ref{ChiCon}) computed for two kinds of
hole-densities: $x=0.5$ and $x=0.3$. 
The parameters used are $g/t^2=0.1$ and $S=3/2$ in both cases.}
\end{center}
\end{figure}
This function plays a crucial role in
the spin-density-wave (SDW) in 1D systems due to the divergence at $p=2k_F$
in the absence of disorder. However, in the system under consideration,
it shows just a peak at $p=2k_F$ due to the self-energy of electrons, i.e,
the scattering of up-spin electrons by the disorder potential. 
In Fig.\ref{fig:Chi}, we show examples of this function
in two kinds of the hole densities 
(of the lower band) $x$ defined by $k_F=\pi(1-x)$.
One can see that the hight of the peak depends weakly on $x$,
since $g$ enters $\chi_p$ through $\theta$. 

\begin{figure}[htb]
\begin{center}
\includegraphics[width=0.5\linewidth]{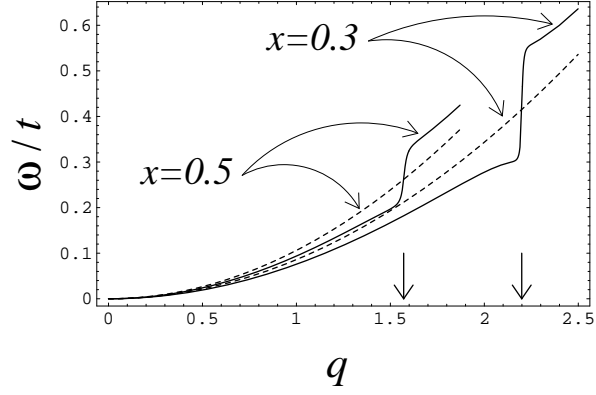}
\caption{\label{fig:MagDis}
The dispersion in eq.(\ref{FinDis}).
The dotted-line denotes the quadratic dispersion in the 
$g=0$ limit (i.e, $\omega_q=t'q^2$).
The down-arrows denote the positions of the Fermi wavenumbers.}
\end{center}
\end{figure}

Having derived the function $\chi_p$, we next calculate the dispersion
of the spin-wave. 
The $p$-integration in eq.(\ref{NexEq}) is rewritten as
\begin{eqnarray}
\sum_p
\frac{\left(p^2q^2+q^4\right)\chi_p^2}{p^2+2pq}
=\frac{1}{(2\pi t)^2}\frac{\tildekf}{4\pi}Q^2
\left[g^{(2)}_Q+\frac{Q^2}{4}g^{(0)}_Q
\right],
\end{eqnarray}
where $Q=q/\tildekf$ and $g^{(n)}_Q$ for $n=0,2$ is defined by
\begin{eqnarray}
g^{(n)}_Q=\frac{1}{2^3}\int_{-\infty}^\infty dx
\left(
\frac{1}{x+Q}+\frac{1}{x-Q}
\right)
\frac{1}{x^{3-n}}
\ln^2\left(
\frac{1+2\cos\frac{\theta}{2}+x^2}{1-2\cos\frac{\theta}{2}+x^2}
\right).
\end{eqnarray}
Using the contour integral in the complex plane,
this integral can be converted into
\begin{eqnarray}
g^{(n)}_Q&=&\frac{\pi\theta\cos\frac{\theta}{2}}{Q^2}\delta_{n,0}
-\frac{\pi}{Q^{3-n}}
\ln
\left|
\frac{e^{-i\frac{\theta}{2}}-Q}{e^{i\frac{\theta}{2}}+Q}
\right|
\Arg
\left(
\frac{e^{-i\frac{\theta}{2}}-Q}{e^{i\frac{\theta}{2}}+Q}
\right)
\nonumber\\
&&
+\pi\int_{\cos\frac{\theta}{2}}^\infty dx\Im
\Bigg[
\left(
\frac{1}{x+Q+i\sin\frac{\theta}{2}}+\frac{1}{x-Q+i\sin\frac{\theta}{2}}
\right)
\frac{1}{(x+i\sin\frac{\theta}{2})^{3-n}}
\nonumber\\
&&~~~\times
\Ln\left(
\frac{\cos\frac{\theta}{2}-2i\sin\frac{\theta}{2}+x}
{\cos\frac{\theta}{2}-2i\sin\frac{\theta}{2}-x}
\right)
\Bigg] .
\label{GFunF}
\end{eqnarray}
This expression is convenient for numerical computation as well as
to take the $g\rightarrow0$ limit, as we shall see momentarily.
The dispersion of the spin-wave is finally given by
\begin{eqnarray}
\omega_q=t'q^2
-\frac{g}{S^2t'}
\frac{1}{(2\pi)^2}\frac{\tildekf}{4\pi}Q^2
\left[g^{(2)}_Q+\frac{Q^2}{4}g^{(0)}_Q
\right] .
\label{FinDis}
\end{eqnarray}
Computing $g^{(n)}_Q$ in eq.(\ref{GFunF}) numerically, 
we show in Fig.\ref{fig:MagDis} examples of the spin-wave dispersion 
for the same systems as those in Fig.\ref{fig:Chi}.
One can indeed find an anomalous dispersion at the 
Fermi wavenumber of electrons. Although gap opening is not observed,
a sharp jump appears just at the Fermi wavenumber. 
Therefore, we can conclude that
the diagram in Fig.\ref{fig:NexSelEne}, which includes
the correlations among up-spin fermions through disorder,
plays an important role in the dynamics of localized spins, even though
the disorder is nonmagnetic.

So far we have derived the spin-wave dispersion in the presence of
disorder by the use of $f_\tau(\varepsilon)$ in the calculation of
$\chi_p$ and hence, of $g_Q^{(n)}$.
To clarify the origin of such  anomalous behavior, let us
take $g\rightarrow0$ limit in eq.(\ref{GFunF}),
as has been done for $\chi_p$ in eq.(\ref{ChiLim}), yielding 
\begin{eqnarray}
g^{(n)}_Q=
-\frac{2\pi^2}{Q^2}\delta_{n,0}\pm
\frac{\pi^2}{Q^{3-n}}\ln\left|\frac{1+Q}{1-Q}\right|
\quad
\mbox{for}\quad\left\{
\begin{array}{l}
|Q|<1\\
|Q|>1
\end{array}\right. .
\end{eqnarray}
If one uses this formula instead of eq.(\ref{GFunF}), 
one reaches the following spin-wave dispersion:
\begin{eqnarray}
\omega_q=(t'+2g\alpha)q^2\mp 5k_Fg\alpha q \ln
\left|\frac{q+k_F}{q-k_F}\right|\quad
\mbox{for}\quad\left\{
\begin{array}{l}
|q|<k_F\\
|q|>k_F
\end{array}\right. ,
\label{MagDisg0}
\end{eqnarray}
where the constant $\alpha$ is given by
$\alpha=1/(64\pi k_F S^2t')$.
The dispersion in eq.(\ref{MagDisg0}) now diverges at $q=k_F$.
Therefore, the present result reveals that 
an enhancement in $\chi_p$ at the nesting vectors 
is responsible for the anomalous gapped-like behavior of the spin-wave
dispersion of localized spins. 
However, it should be stressed that even though the divergence in 
$\chi_p$ is suppressed by disorder, which is due to the smearing of the 
Fermi surface, singular behavior of spin-wave dispersion seems robust.

So far we have demonstrated that nonmagnetic impurities of
conduction electrons actually induce the anomalous behavior of localized 
spins at the Fermi wavenumbers of electrons, as suggested by
Motome and Furukawa, by using the leading order perturbation theory.
In models with small $J_H$,
the diagrams in Fig.\ref{fig:LedSelEne} leads to interesting
physics such as SDW or spin glass 
(RKKY exchange interaction). Contrary to these,
the same diagram merely yields the canonical
dispersion of the spin-wave in the present systems
with large $J_H$ because of weak correlations between
electrons with different spins.
Even in such systems, however,
disorder induces correlations among up-spin electrons, and 
strongly affects the magnetic properties, 
even if disorder is weak and nonmagnetic.

Finally, we would like to discuss the models in higher dimensions.
The chief ingredient in a step-function-like behavior in the dispersion
is the function $\chi_p$, which has a sharp peak at $p=2k_F$ in 1D. 
This reflects the nesting of the Fermi surface.
In higher dimensions, however, this function is
continuous if we use the free electron (quadratic) dispersion. 
Therefore, in such an approximation,
one cannot expect any anomalous dispersions in the spin-wave 
excitation. 
On the other hand, 
the tight-binding electron dispersion gives rise to an enhancement
around the nesting wavevectors of electrons near half-filling.
Therefore, one can expect  similar anomalous behavior of
spin-waves there even in higher dimensions.
The question is how robust such anomalous behavior is away from half-filling.
Unfortunately, cosine dispersion of electrons
does not allow analytic calculations in higher dimensions,
and therefore, the details, including numerical results, will 
be published elsewhere.

The authors would like to thank 
J. Igarashi, Y. Motome and M. Yokoyama 
for valuable discussions.
This work was supported in part by Grant-in-Aid for Scientific Research
from JSPS.



\begin{thebibliography}{9}
%
\bibitem{HWH}
R. von Helmolt,
 J. Wecker, B. Holzapfel, L. Schultz and K. Samwer:
Phys. Rev. Lett. {\bf 71} (1993) 2331.
%
\bibitem{COK}
K. Chahara, T. Ono, M. Kasai and Y. Kozono:
Appl. Phys. Lett. {\bf 63} (1993) 1990.
%
\bibitem{JTM}
S. Jin, T. H. Tiefel, M. McCormack, R. A. Fastnacht, R. Ramesh and L. H. Chen:
Science {\bf 264} (1994) 413.
%
\bibitem{IFT}
M. Imada, A. Fujimori and Y. Tokura:
Rev. Mod. Phys. {\bf 1998} (1998) 1039.
\bibitem{SalJai}
M. B. Salamon and M. Jaime:
Rev. Mod. Phys. {\bf 73} (2001) 583:
%
\bibitem{DHM}
E. Dagotto, T. Hotta and A. Moreo:
Phys. Rep. {\bf 344} (2001) 1.
%

\bibitem{Zen} 
C. Zener:
Phys. Rev. {\bf 82}, 403 (1951).
%
\bibitem{AndHas}
P. W. Anderson and H. Hasegawa:
Phys. Rev. {\bf 100}, 675 (1995).
%
\bibitem{Gen}
P. G. de Gennes:
Phys. Rev. {\bf 118}, 141 (1960).
%
\bibitem{KubOha}
K. Kubo and N. Ohata:
J. Phys. Soc. Jpn. {\bf 33}, 21 (1972).
%
%
\bibitem{HDC}
H. Y. Hwang, P. Dai, S-W. Cheong, G. Aeppli, D. A. Tennant and H. A. Mook:
Phys. Rev. Lett. {\bf 80} (1998) 1316.
%
\bibitem{DLM}
L. Vasiliu-Doloc, J. W. Lynn, A. H. Moudden, A. M. de Leon-Guevara and A. Revcolevschi:
Phys. Rev. {\bf B58} (1998) 14913.
%
\bibitem{DHZ}
P. Dai, H. Y. Hwang, Jiandi Zhang, J. A. Fernandez-Baca,
S.-W. Cheong, C. Kloc, Y. Tomioka and Y. Tokura:
Phys. Rev. {\bf B61} (2000) 9553.
%
\bibitem{BHM}
G. Biotteau, M. Hennion, F. Moussa, J. Rodr\'iguez-Carvajal, 
L. Pinsard, A. Revcolevschi, Y. M. Mukovskii and D. Shulyatev:
Phys. Rev. {\bf B64} (2001) 104421.
%
\bibitem{AUT}
D. Akahoshi, M. Uchida, Y. Tomioka, T. Arima, Y.Matsui and Y. Tokura:
Phys. Rev. Lett. {\bf 90} (2003) 177203.

%
\bibitem{Var}
C. M. Varma:
Phys. Rev. {\bf B54} (1996) 7328.
%
\bibitem{MueDag}
E. M\"uller-Hartmann and E. Dagotto:
Phys. Rev. {\bf B54} (1996) 6819.
%
\bibitem{SXS1}
L. Sheng, D. Y. Xing, D. N. Sheng and C. S. Ting:
Phys. Rev. Lett. {\bf 79} (1997) 1710.
%
\bibitem{SXS2}
L. Sheng, D. Y. Xing, D. N. Sheng and C. S. Ting:
Phys. Rev. {\bf B56} (1997) 7053.
%
\bibitem{AllAla}
R. Allub and B. Alascio:
Phys. Rev. {\bf B55} (1997) 14113.
%
\bibitem{BMM}
J. Burgy, M. Mayr, V. Martin-Mayor, A. Moreo and E. Dagotto:
Phys. Rev. Lett. {\bf 87} (2001) 277202.
%
\bibitem{Fur99}
N. Furukawa:
J. Phys. Soc. Jpn. {\bf 68} (1999) 2522.
%
\bibitem{KhaKil}
G. Khaliullin and R. Kilian:
Phys. Rev. {\bf B61} (2000) 3494.
%
\bibitem{SolTer}
I. V. Solovyev and K. Terakura:
Phys. Rev. Lett. {\bf 82} (1999) 2959.
%
\bibitem{Gol}
D. I. Golosov:
Phys. Rev. Lett. {\bf 84} (2000) 3974.
%
\bibitem{ShaChu}
N. Shannon and A. V. Chubukov:
Phys. Rev. {\bf B65} (2002) 104418.
%
%
\bibitem{MotFur1}
Y. Motome and N. Furukawa:
J. Phys. Soc. Jpn. {\bf 71} (2002) 1419.
%
\bibitem{MotFur2}
Y. Motome and N. Furukawa:
J. Phys. Soc. Jpn. {\bf 72} (2003) 472.
%
\bibitem{MotFur3}
Y. Motome and N. Furukawa:
preprint, cond-mat/0305488.
%
\bibitem{Fur96}
N. Furukawa:
J. Phys. Soc. Jpn. {\bf 65} (1996) 1174.
\end{thebibliography}
\end{document}